\documentclass{ifacconf}
\usepackage{graphicx}
\usepackage{float}
\usepackage{epsf}
\usepackage{amssymb}
\usepackage{fancyhdr}
\usepackage{subfigure}
\usepackage{latexsym}
\usepackage{amsmath}
\usepackage{natbib}

\newcommand{\var} {\mbox{\rm var}}

\newcommand{\virg}[1]{``#1''}

\newcommand{\Ga} {\mbox{\rm Ga}}

\newtheorem{theorem}{Theorem}[section]
\newtheorem{lemma}[theorem]{Lemma}

\newtheorem{remark}[theorem]{Remark}

\begin{document}

\begin{frontmatter}

\title{Bayesian kernel-based system identification with quantized output data} 

\thanks[footnoteinfo]{The research leading to these results has received funding
from the Swedish Research Council under contract 621-2009-4017 and
the European Union Seventh Framework Programme
[FP7/2007-2013] under grant agreement no. 257462 HYCON2 Network of excellence, by the MIUR
FIRB project RBFR12M3AC - Learning meets time:
a new computational approach to learning in dynamic
systems}

\author[First]{Giulio Bottegal}
\author[Second]{Gianluigi Pillonetto}
\author[First]{H\r akan Hjalmarsson}

\address[First]{ACCESS Linnaeus Centre, School of Electrical Engineering,
KTH Royal Institute of Technology, Stockholm, Sweden \\(e-mail: \{bottegal; hjalmars\}@kth.se)}
\address[Second]{Department of Information Engineering, University of Padova, Padova, Italy  (e-mail: giapi@dei.unipd.it)}

\begin{abstract}
In this paper we introduce a novel method for linear system identification with quantized output data. We model the impulse response as a zero-mean Gaussian process whose covariance (kernel)
is given by the recently proposed stable spline kernel, which encodes information on regularity and exponential stability. This serves as a starting point to cast our system identification problem into a Bayesian framework. We employ Markov Chain Monte Carlo (MCMC) methods to provide an estimate of the system. In particular, we show how to design a Gibbs sampler which quickly converges to the target distribution. Numerical simulations show a substantial improvement in the accuracy of the estimates over state-of-the-art kernel-based methods when employed in identification of systems with quantized data.
\end{abstract}

\end{frontmatter}

\section{Introduction}
Identification of systems from quantized data finds applications in a wide range of areas such as communications, networked control systems, bioinformatics (see e.g. \citep{bae2004gene} and \citep{wang2010system}).

From a system identification perspective, identification of systems having quantized output data constitutes a challenging problem. In fact, the presence of a quantizer cascaded to a dynamic system, causes a loss of information on the behavior of that dynamic system. Thus, standard system identification techniques, such as least-squares or prediction error method (PEM) \citep{Ljung}, \citep{Soderstrom}, may give poor performances. For this reason, in recent years several techniques for system identification from quantized data have been proposed in a series of papers. Some of these methods are specifically tailored for identification of systems with binary measurements \citep{wang2003system},
 \citep{wang2006joint}, and are possibly implemented in a recursive fashion
\citep{guo2013recursive}, \citep{jafari2012convergence}. Other methods, such as \citep{colinet2010weighted}, exploit the knowledge of a dithering signal to improve the identification performances. Specific input design techniques are studied in \citep{godoy2014novel}, \citep{casini2011input} and \citep{casini2012input}. Methods for handling general types of quantization of data have been proposed recently  \citep{godoy2011identification}, \citep{chen2012impulse}. In such contributions, the problem of identifying a linear dynamic system with quantized data is posed as a likelihood problem. In particular, in \citep{chen2012impulse}  authors  exploit the recently proposed Bayesian kernel-based formulation of the linear dynamic system identification problem (see \citep{pillonetto2014kernel} for a survey).

Similarly to \citep{chen2012impulse}, the starting point of this paper is the formulation of the problem of identifying a linear dynamic systems with quantized data using a Bayesian approach. We model the impulse response of the unknown system as a realization of a Gaussian random process. Such a process has zero mean and its covariance matrix (in this context also called a \emph{kernel}) is given by the recently introduced \emph{stable spline kernels}, \citep{SS2010}, \citep{SS2011}, \citep{bottegal2013regularized} which are specifically designed for linear system identification purposes. The structure of this type of kernels depends on two \emph{hyperparameters}, namely a \emph{scaling} parameter and a \emph{shaping} parameter, whose tuning permits more flexibility in the identification process and can be seen as a model selection step. In the standard setting (i.e. when there is no quantizer), kernel hyperparameters are chosen as those maximizing the marginal likelihood of the output measurements, obtained by integrating out the dependence on the system. Once the hyperparameters are chosen, the impulse response of the system is computed as the minimum mean-square Bayes estimate given the observed input/output data (see e.g. \citep{SS2010}, \citep{ChenOL12}).

A key assumption in kernel-based methods is that the output data and the system admit a joint Gaussian description. Such an assumption does not hold with quantized data and we need to think of a different approach. In this paper we propose a solution based on Markov Chain Monte Carlo (MCMC) techniques \citep{Gilks}. To this end, we define a target probability density; the estimate of the system can be obtained by drawing samples from it. Such a probability density is function of the following random variables: $1)$ the (unavailable) non-quantized output of the linear system, $2)$ the scaling hyperparameter of the kernel, $3)$ the unknown measurement noise variance, $4)$ the impulse response of the system. The main contribution of this paper is to show how to design a Gibbs sampler \citep{Gilks} by exploiting the knowledge of the conditional densities of the target distribution. The main advantage of the Gibbs sampler is that it does not require any rejection criterion of the generated samples and quickly converges to the target distribution.
Note that MCMC-based approaches have recently gained popularity in system identification \citep{Ninness2010}, \citep{lindsten2012}, \citep{bottegal2014outlier}.

The paper is organized as follows. In the next section, we introduce the problem of the identification of dynamic systems from quantized data. In Section \ref{sec:Bayesian}, we give a Bayesian description of the variables entering the system. In Section \ref{sec:proposed_method}, we describe the proposed method for identification. Section \ref{sec:experiments} shows some simulations to assess the performances of the proposed method. Some conclusions end the paper.

\section{Problem statement}
We consider the following linear time-invariant BIBO stable output error system
\begin{equation} \label{eq:oe_model}
z_t = (g \ast u)_t + v_t \,,
\end{equation}
where $\{g_t\}, \,t \in \mathcal{T}$ is the impulse response characterizing the unknown system, which is fed by the input $\{u_t\}, \,t \in \mathcal{T}$. The set $\mathcal{T}$ corresponds to either $\mathbb{R}^+$ or $\mathbb{Z}^+$, depending on whether the system is continuous-time or discrete-time. The output $z_t$ is corrupted by the additive white Gaussian noise $v_t$, which has zero mean and unknown variance $\sigma^2$, and measured at the time instants $t \in \mathcal{I}$. If the system is continuous-time, then $\mathcal{I}$ can represent any non-uniform sampling, whereas in the discrete-time case we shall consider $\mathcal{I} \equiv \mathbb{Z}^+$ (i.e., no downsampling). For ease of exposition, in this paper we shall derive our algorithm in the discrete-time case only; the extension to the continuous-time is quite straightforward (see e.g. \citep{Wahba1990}, \citep{SS2010}).
\begin{figure}[!ht]
\begin{center}
{\includegraphics[width=7.5cm]{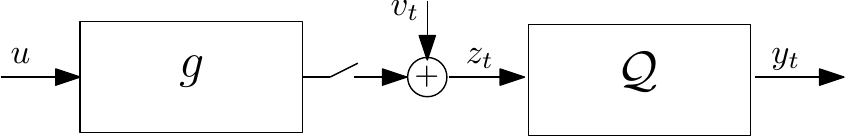}}
 \caption{\emph{Block scheme of the system identification scenario.}} \label{fig:block_scheme}
\end{center}
\end{figure}
Actually, the output $z_t$ is not directly  measurable, and only a quantized version is available, namely
\begin{equation}
y_t = \mathcal Q [z_t] \,,
\end{equation}
where $\mathcal Q$ is a known map (our quantizer) of the type
\begin{equation}
\mathcal Q[x] = p_k \qquad \mbox{if } x \in (q_{k-1},\,q_k] \,,
\end{equation}
with $p_k \in \{p_1,\,\ldots,\,p_Q\}$ and $q_k \in \{q_0,\,\ldots,\,q_Q\}$ being known (and typically $q_0 = -\infty$ and $q_Q = \infty$).

\begin{remark}
A particular and well-studied case is the binary quantizer, defined as
\begin{equation}
\mathcal Q[x] = \left\{ \begin{array}{ll} -1 & \mbox{if } x < C \\
								  1 & \mbox{if } x \geq C
		\end{array}  \right. \,.
\end{equation}
It is well-known that a condition on the threshold to guarantee identifiability of the system is $C \neq 0$. In fact, when $C=0$, the system can be determined up to a scaling factor \citep{godoy2011identification}.
\end{remark}

Without loss of generality, let us assume the system to be strictly causal, i.e. $g_0 = 0$. We assume that $N$ input-output data samples $y_1,\,\ldots,\,y_N$, $u_0,\,\ldots,\,u_{N-1}$ are collected during an experiment. We formulate our system identification problem as the problem of estimating the impulse response $g$ for $n$ time instants, namely obtain  $\{g_t\}_{t=1}^n$. Recall that, if $n$ is sufficiently large, these samples can be used to approximate the dynamics of the systems with arbitrary accuracy \citep{ljung1992}.
Introducing the vector notation
$$
g := \begin{bmatrix} g_1 \\ \vdots \\ g_n \end{bmatrix} ,\, y := \begin{bmatrix} y_1 \\ \vdots \\ y_N \end{bmatrix} \,,\, z := \begin{bmatrix} z_1 \\ \vdots \\ z_N \end{bmatrix} \,,\,  v := \begin{bmatrix} v_1 \\ \vdots \\ v_N \end{bmatrix}
$$
$$
U = \begin{bmatrix} u_0 & 0  & & \ldots & 0 \\ u_1 & u_0 & 0 & \ldots & 0 \\ \vdots & \vdots &  &\ddots & \vdots \\ u_{N-2} & u_{N-3}  &  \ldots & u_{N-n+1}& 0 \\ u_{N-1} & u_{N-2}  &  \ldots &\ldots &  u_{N-n}  \end{bmatrix} \, \in \, \mathbb{R}^{N \times n}  \,,
$$
the input-output relation for the available samples can be written
\begin{align*} \label{eq:sys2}
z & = Ug + v  \nonumber \\
y_t & = \mathcal Q [z_t] \quad,\, t = 1,\, \ldots,\,N
\end{align*}
so that our estimation problem can be cast in, say, a \virg{linear regression plus quantization} form.

\section{Bayesian models for the quantities of interest} \label{sec:Bayesian}

\subsection{Establishing a prior for the system}

In this paper we cast the system identification problem into a Bayesian framework. Our starting point is the setting of a proper prior on $g$. Following a Gaussian regression approach \citep{Rasmussen}, we model $g$ as a zero-mean Gaussian random vector, i.e. we assume the following probability density function for $g$:
\begin{equation}\label{eq:model_g}
p(g|\lambda,\,\beta) \sim \mathcal N (0,\,\lambda K_{\beta}) \,,
\end{equation}
where $K_\beta$ is a covariance matrix whose structure depends on the value of the \emph{shaping hyperparameter} $\beta$ and $\lambda \geq 0$ is the \emph{scaling hyperparameter}.
In this context, $K_\beta$ is usually called a {\it kernel}  and determines the properties of the realizations of $g$.
In this paper, we choose $K_\beta$ from the family of \emph{stable spline kernels} \citep{SS2010}, \citep{SS2011}. Such kernels are specifically designed for system identification purposes and give clear advantages compared to other standard kernels \citep{bottegal2013regularized}, \citep{SS2010} (like the quadratic kernel or the Laplacian kernel, see \citep{Scholkopf01b}).
In this paper we make use of the \emph{first-order stable spline kernel} (or \emph{TC kernel} in \citep{ChenOL12}). It is defined by
\begin{equation} \label{eq:ssk1}
\{K_\beta\}_{i,j} := \beta^{ \max(i,j)} \quad,\, 0 < \beta  <1 \,,
\end{equation}
The above kernel is parameterized by $\beta$, which regulates the decaying velocity of the generated impulse responses. 

\subsection{Bayesian description of the non-quantized output} \label{sec:non-quantized}

Since we have assumed Gaussian distribution of the noise $v$, the joint distribution of the vectors $z$ and $g$, given values of $\lambda$, $\beta$ and the noise variance $\sigma^2$, is jointly Gaussian, namely
\begin{equation} \label{eq:joint_Gaussian}
p\left(\left.\begin{bmatrix} z \\ g \end{bmatrix}\right|\lambda,\,\beta,\,\sigma^2 \right) \sim \mathcal N \left( \begin{bmatrix} 0\\0 \end{bmatrix} , \begin{bmatrix} \Sigma_z & \Sigma_{zg} \\ \Sigma_{gz} & \lambda K_\beta \end{bmatrix} \right)\,,
\end{equation}
where
\begin{equation} \label{eq:var_z}
\Sigma_z = \lambda U K_\beta U^T + \sigma^2 I_N
\end{equation}
and $\Sigma_{zg} = \Sigma_{gz}^T =  \lambda U K_\beta$. It follows also that the posterior distribution of $g$, given the knowledge of $z$, is Gaussian, namely
\begin{equation} \label{eq:pg}
p(g|z,\,\lambda,\,\beta,\,\sigma^2) = \mathcal N\left(C z,\,P \right) \,,
\end{equation}
where
\begin{equation} \label{eq:CandP}
P  = \left( \frac{U^T U}{\sigma^2} + (\lambda K_\beta)^{-1} \right)^{-1} \quad,\quad C  = P \frac{U^T}{\sigma^2}  \,.
\end{equation}
In \citep{SS2010}, an impulse response estimator is derived starting from \eqref{eq:pg}. In fact, the minimum mean-squared error (MMSE) estimate of $g$ is (see e.g. \citep{Anderson:1979})
\begin{equation} \label{eq:est_g}
\hat g = \mathbb{E} [g|z,\,\lambda,\,\beta,\,\sigma^2] = Cz \,.
\end{equation}
Such an estimator depends on the kernel hyperparameters and the noise variance. A common strategy to choose the kernel hyperparameters is to maximize the marginal likelihood of $z$, that is
\begin{align} \label{eq:marg_lik}
(\hat \lambda,\,\hat \beta) & = \arg \max_{\lambda,\,\beta} \log p(z|\lambda,\,\beta) \nonumber \\
& = \arg \min_{\lambda,\,\beta} \log\det \Sigma_z + z^T \Sigma_z^{-1} z \,.
\end{align}
An estimate $\hat \sigma^2$ of $\sigma^2$ can be computed by means of the following steps:
\begin{enumerate}
\item Compute the least-squares estimate of $g$, i.e.
\begin{equation}
\hat g_{LS} = (U^T U)^{-1}U^T z \,,
\end{equation}
in order to obtain an estimate of $g$;
\item Compute the empirical estimate of $\sigma^2$
\begin{equation} \label{eq:hat_sigma}
\hat \sigma^2 =  \frac{\left(z-U\hat g_{LS}  \right)^{T}\left(z-U\hat g_{LS} \right)}{N-n} \,.
\end{equation}
\end{enumerate}

Clearly, the system identification method described above is not applicable in our problem, since $z$ is not available. However, we can draw some information on such a vector from the quantized output $y$. First, note that from \eqref{eq:joint_Gaussian} it follows that
\begin{equation} \label{eq:pz}
p(z|g,\,\sigma^2) = \mathcal N\left(U g,\,\sigma^2 I \right) \,.
\end{equation}
Note also that, once $g$ is given, \eqref{eq:pz} is independent of $\lambda$ and $\beta$.
Let $U_t$ denote the $t$-th row of $U$. Then, for each entry $z_t$ it holds that (see e.g. \citep{bae2004gene})
\begin{equation} \label{eq:z_i}
p(z_t| g,\, \sigma^2 ,\, y_t = p_k) = \mathcal N_{q_{k-1}}^{q_k} (U_t g,\,\sigma^2) \,,
\end{equation}
where $\mathcal N_{a}^{b}(\mu,\,\sigma^2)$ denotes a Gaussian distribution truncated below $a$ and above $b$, whose original mean and variance are $\mu$ and $\sigma^2$ respectively. Note that, for $t \neq j$
\begin{align}
p(z_t,\,z_j| g ,\, \sigma^2 ,\, y)  = p(z_t| g ,\, \sigma^2 ,\, y) p(z_j| g ,\, \sigma^2 ,\, y)\,,
\end{align}
due to the assumption on whiteness of noise.

\subsection{Bayesian description of hyperparameters and noise variance}

As mentioned in the previous subsection, knowing the values of hyperparameters is of paramount importance in kernel-based methods. The marginal likelihood maximization approach \eqref{eq:marg_lik} is not applicable here, so we have to think of alternative ways to estimate the values of the hyperparameters.

Let us denote by $\Ga(a,\,b)$ the Gamma distribution with parameters $a$ and $b$.
The following result, drawn from \citep{MagniPAMI}, shows the marginal distribution of the inverse of the hyperparameter $\lambda$ given $g$ and $\beta$.

\begin{lemma}\label{lem:lambda}
The posterior probability distribution of $\lambda^{-1}$ given $g$ and $\beta$ is
\begin{equation} \label{eq:Gamma_dist}
p(\lambda^{-1}|g,\,\beta) \sim \Ga \left(\frac{n}{2},\,\frac{g^T K_\beta^{-1} g}{2} \right)
\end{equation}
\end{lemma}
\begin{remark}
To obtain the result of the above lemma, we have implicitly set an improper prior on $\lambda$ with non-negative support, i.e., $p(\lambda) = \lambda^{-1} \chi_+(\lambda)$, where $\chi_+(\lambda)$ is the indicator function with support $\mathbb{R}^+$.  
\end{remark}

A similar argument holds for the noise variance $\sigma^2$. Recalling that $v$ is Gaussian with variance  equal to $\sigma^{2}I$ and readapting Lemma \ref{lem:lambda} to this case, it turns out that
\begin{equation} \label{eq:Gamma_dist_sigma}
p(\sigma^{-2}|v) \sim \Ga \left(\frac{N}{2},\,\frac{v^T v}{2}\right) \,,
\end{equation}
where also here we have assumed the improper prior $p(\sigma^2) = \sigma^{-2} \chi_+(\sigma^2)$.

In some situations, e.g. when the quantizer has mild effects on the measured signal (e.g., when the resolution of the quantizer is high), a sufficiently reliable estimate of $\sigma^2$  can be obtained by using \eqref{eq:hat_sigma}, with $z$ replaced by $y$.
However, note that  in general $\hat \sigma^2$ is not a consistent estimate of $\sigma^2$.

We conclude this section by recalling that, unfortunately, establishing a Bayesian model for  $\beta$ is still an unsolved problem. In this paper, we shall consider such an hyperparameter as deterministic. A method for its choice will be discussed in the next section.

%
%


\section{Proposed system identification method} \label{sec:proposed_method}

In this section we show how to exploit the Bayesian models introduced in the previous section to derive our system identification method.
Assume for the moment that $\beta$ is known and let us drop it from the notation below. The impulse response estimate $\hat g$ can be obtained by computing the following integral
\begin{equation} \label{eq:bayes_est3}
\hat g = \int g\; p(z,\,\lambda,\,\sigma^2,\,g|y)\; dz\; d\lambda \; d\sigma^2 \;  dg \,,
\end{equation}
where
\begin{equation} \label{eq:target_distr}
p(z,\,\lambda,\,\sigma^2,\,g|y)
\end{equation}
denotes the joint distribution of the random quantities described in the previous section, given the quantized output $y$. The above integral can be computed using Markov Chain Monte Carlo (MCMC) methods \citep{Gilks}, by drawing a large number of samples from the  distribution $p(z,\,\lambda,\,\sigma^2,\,g|y)$ (usually called \emph{target distribution}) and then averaging over $g$. In general, drawing samples from a distribution is a hard problem, if its probability density function does not admit a closed-form expression. However, if all the conditional probability densities of such a distribution are available in closed-form, the problem of sampling from the target distribution can be solved efficiently by resorting on a special case of the Metropolis-Hastings method, namely the Gibbs sampler (see e.g. \citep{Gilks}).
The basic idea is that each conditional random variable is the state of a Markov chain; then, by drawing samples from each conditional probability density iteratively,
we converge to the stationary state of this Markov chain and generate samples of the target distribution. Here, the conditionals of \eqref{eq:target_distr} are as follows.

\begin{enumerate}
\item $p(z|\lambda,\,\sigma^2,\,g,\,y)$. As discussed in Section \ref{sec:non-quantized}, once $g$ is given, this conditional density does not depend on $\lambda$. Moreover, due to the assumptions on noise, it density factorizes as follows
\begin{equation}
\prod_{t=1}^N p(z_t|   \sigma^2 ,\,g,\, y_t) \,,
\end{equation}
where each of the factors is a truncated Gaussian according to \eqref{eq:z_i}.
\item $p(\lambda^{-1}|z,\,\sigma^2,\,g,\,y)$. Once $g$ is given, $\lambda$ becomes independent of all the other variables, i.e. such conditional density becomes
$
p(\lambda^{-1}|g) \,,
$
and corresponds to
\eqref{eq:Gamma_dist}, namely a Gamma distribution with parameters $(\frac{n}{2},\,\frac{g^T K_\beta^{-1} g}{2})$.
\item $p(\sigma^2|z,\,\lambda,\,g,\,y)$. Once $g$ and $z$ are given, one can compute
$
v = z - Ug \,.
$
Recalling \eqref{eq:Gamma_dist_sigma}, it follows that this conditional density can be written as
$
p(\sigma^2|z,\,g)
$
and is distributed as a Gamma random variable with parameters $(\frac{N}{2},\,\frac{(z- Ug)^T(z-Ug)}{2})$.
\item $p(g|z,\,\lambda,\,\sigma^2,\,y)$. Given $z$, information carried by $y$ becomes redundant and can be discarded, so that this conditional probability density corresponds to
$
p(g|z,\,\lambda,\,\sigma^2) \,.
$
Its closed-form expression is given by \eqref{eq:pg}, namely a Gaussian distribution with mean $C z$ and covariance matrix $P$ (see \eqref{eq:CandP}).
\end{enumerate}

Given the above conditional densities, we are in position to illustrate the proposed identification algorithm.
\begin{algorithm}[ht!] \label{alg}
\textbf{Algorithm}: Bayesian system identification with quantized output measurements \vspace{0.1cm}\\
Input: $\{y_t\}_{t=1}^N,\,\{u_t\}_{t=0}^{N-1}$ \vspace{0.1cm} \\
Output: $\{\hat{g}\}_{t=1}^n$
\begin{enumerate}
\item Initialization: Compute initial values $g^0$, $\sigma^{2,0}$ and set $\beta$
\item For $i =1$ to $M$:
            \begin{enumerate}
                \item  Draw the sample $z_t^i$, from $p(z_t|g^{i-1},\,\sigma^{2,i-1},\,y_t)$, $t=1,\,\ldots,\,N$  %
                \item  Draw the sample $\lambda^i$ from $p(\lambda^{-1}|g^{i-1})$ 
				\item Draw the sample $\sigma^{2,i}$ from $p(\sigma^2|g^{i-1},\,z^{i}) $
                \item Draw the sample $g^i$ from $p(g|\lambda^{i},\,z^{i},\,\sigma^{2,i})$
            \end{enumerate}
\item Compute $\hat g = \frac{1}{M-M_0} \sum_{i=M_0}^M g^i$
\end{enumerate}
\end{algorithm}

In the  above algorithm, the parameters $M$ and $M_0$ are  introduced. $M$ the number of samples to be generated; clearly, large values of $M$ should guarantee more accurate estimates of $g$. $M_0$ is the number of initial samples drawn from the conditional of $g$ to be discarded and is also known as \emph{burn-in} \citep{meyn2009markov}. In fact, the conditionals from which those samples are drawn are to be considered as non-stationary, since the Gibbs sampler takes a certain number of iterations to get close to a stationary distribution.

\subsubsection*{Setting initial values of the Gibbs sampler}
The initial estimate $g^0$ can be computed using the kernel-based method introduced in \citep{SS2010} and briefly revisited in Section \ref{sec:non-quantized}. Replacing $z$ with $y$ in \eqref{eq:est_g}, one can obtain a (very) rough estimate of $g$, which can serve as initial condition for the Gibbs sampler.

Similarly, the initial value $\sigma^{2,0}$ can be computed from \eqref{eq:hat_sigma}, again  by replacing $z$ with $y$.

\subsection{Estimation of the hyperparameter $\beta$}

It remains to set a scheme for estimating $\beta$. In \citep{chen2012impulse}, an exact marginal likelihood maximization approach is proposed, but it is shown that such an approach needs a solution of a complicated integral. In this paper, we adopy a simple (and approximate) way to estimate $\beta$. It consists in maximizing the cost function of \eqref{eq:marg_lik}, where, instead of using the non-available data $z$, we plug the quantized output $y$. Clearly, one should not expect to get very good results in general, especially when the difference between $z$ and $y$ is high (e.g. when the quantizer is binary). However, numerical experiments have shown that the accuracy on the estimation of $\beta$ with this strategy is satisfactory enough in order to obtain a good performance of the algorithm. 
\subsection{Comparison with \citep{chen2012impulse}}
Although the methods proposed here and the method proposed in \citep{chen2012impulse} exploit the same Bayesian modeling of the unknown system, the techniques used to carry out the estimate of $g$ are substantially different. In \citep{chen2012impulse}, the impulse response is obtained as the maximum a posteriori (MAP) estimate given the quantized output data. Here instead, $g$ is computed by means of \eqref{eq:bayes_est3}, that is a minimum mean square error Bayes estimator.

\section{Numerical experiments} \label{sec:experiments}

We test the proposed algorithm by means of 2 Monte Carlo experiments of 100 runs each. For each Monte Carlo run, we generate a linear system by picking 10 pairs of complex conjugate zeros with magnitude randomly chosen in $[0,\,0.95]$ and random phase. Similarly, we pick 10 pairs of complex conjugate poles with magnitude randomly chosen in $[0,\,0.93]$ and random phase. The goal is to estimate  $n=50$ samples of the impulse response from $N$ input-output data. The inputs are realizations of white noise with unit variance. We compare the following estimators.
\begin{itemize}
\item B-Q-GS: This is the method described in this paper, namely a Bayesian system identification method for Quantized output data that uses the Gibbs Sampler. The parameter $M$, denoting the number of samples generated by the sampler, is set to $3 \cdot 10^3$. The first $M_0 = 10^3$ samples are discarded. The validity of the choice of $M$ and $M_0$ is checked  by assessing that quantiles 0.25, 0.5, 0.75 are estimated with good precision \citep{Raftery1996}. 
\item SS-ML: This is the nonparametric kernel-based method proposed in \citep{SS2010} and revisited in \citep{ChenOL12}, which is not designed for handling quantized data. This method requires the estimation of the same parameters as our proposed method. The kernel adopted for identification is \eqref{eq:ssk1}. Its hyperparameters are estimated using \eqref{eq:marg_lik}, while $\sigma^2$ is estimated through \eqref{eq:hat_sigma} (in both cases replacing $z$ with $y$).
\item LS: This is the least-squares estimator, where the data employed to estimate $g$ are the quantized output measurements $y$. Note that, in principle, here the parameter $n$ should be estimated from data using complexity criteria such as AIC or BIC \citep{Ljung}. Here, for simplicity, we fix it to 50.
\item SS-ML-NQ: Same as SS-ML. However, here we make use of the non-quantized vector $z$. Hence, this estimator exploits information which is not available in practice in this problem.
\item LS-NQ: Least-squares estimator having access to the vector $z$. The parameter $n$ is fixed as for the LS estimator.
\end{itemize}
The performances of the estimators are evaluated by means of the fitting score, computed as
\begin{equation}
FIT_i  = 1-\frac{\|g_i - \hat g_i \|_2}{\|g_i - \bar g_i\|_2} \,,
\end{equation}
where $g_i$ is the impulse response generated at the $i$-th run, $\bar g_i$ its mean and $\hat g_i$ the estimate computed by the tested methods.

\subsection{Binary quantizer}

\begin{figure*}[!ht]
\begin{center}
\begin{tabular}{cc}
    {\includegraphics[width=7cm]{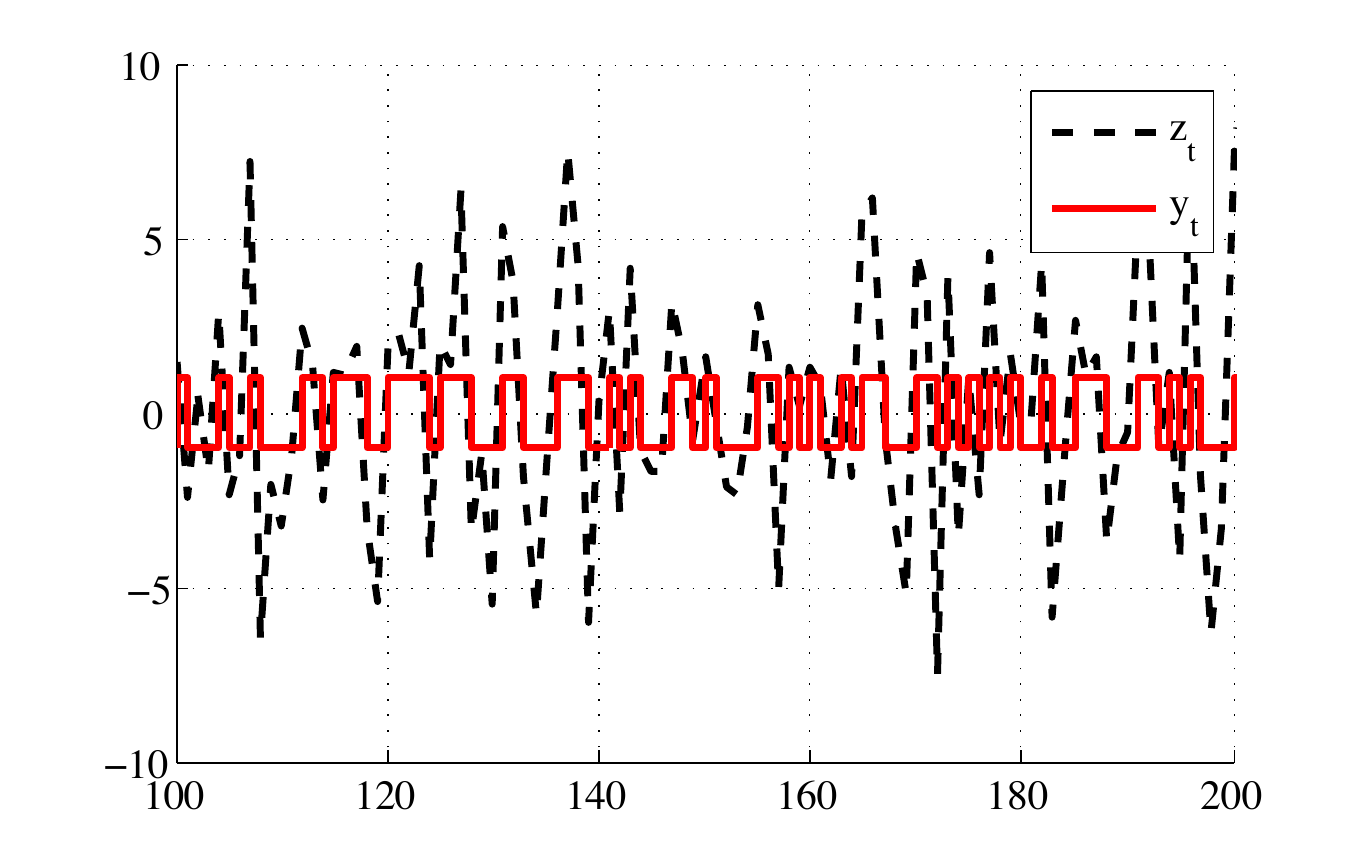}}
&
    {\includegraphics[width=7cm]{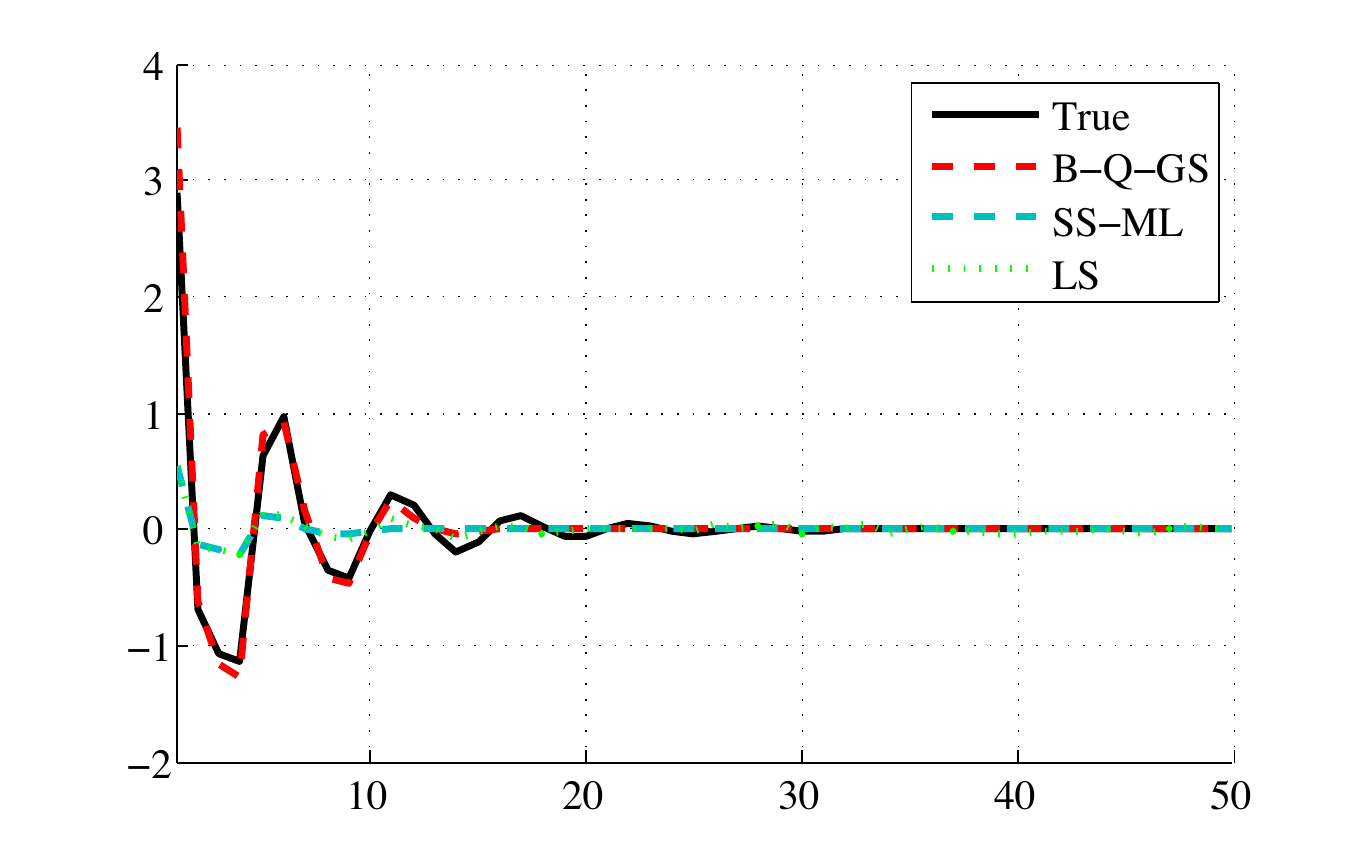}} \\
\end{tabular}
    \caption{\emph{Left: Example of output data with the binary quantizer (samples between time instants 100 and 200). Right: Impulse response of the system generating the data and its estimates.}} \label{fig:example_bin}
\end{center}
\end{figure*}

The first experiment is on the following binary quantizer
$$
\mathcal Q_b [x] := \left\{ \begin{array}{ll} 1 & \mbox{ if } x \geq 1 \\ -1 & \mbox{ if } x < 1 \end{array} \right.\,.
$$
For each Monte Carlo run, the noise variance is such that $\frac{\var (Ug) }{\sigma^2} = 10$, i.e. the ratio between the variance of the noiseless (non-quantized) output and the noise is equal to 10. We generate $N = 500$ data samples.
\begin{figure}[!ht]
\begin{center}
    {\includegraphics[width=7cm]{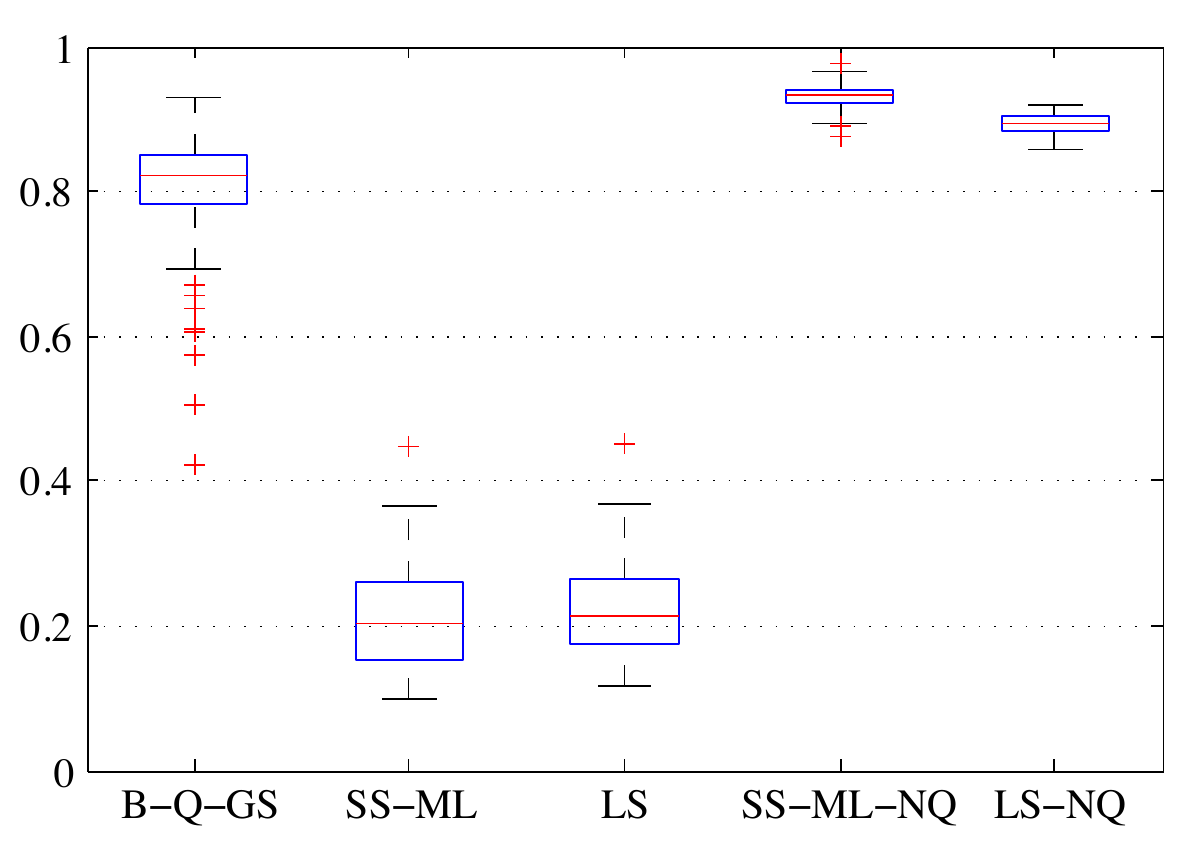}}
\caption{\emph{Box plots of the fitting scores for the binary quantizer experiment.}} \label{fig:bin_exp}
\end{center}
\end{figure}
Figure \ref{fig:bin_exp} shows the results of the Monte Carlo runs. The advantage of using the proposed identification technique, compared to methods which do not account for the quantizer, is evident. Despite the large loss of information caused by the quantizer, the proposed method gives a fit which is quite comparable to the oracle methods. Figure \ref{fig:example_bin} reports one of the generated scenarios. It can be seen that there is a substantial difference between $y$ and $z$. Nonetheless, the accuracy of the estimation of the impulse response is acceptable.

\subsection{Ceil-type quantizer}

In the second experiment we test the performance of our method on systems followed by a ceil-type quantizer, which is defined as
$$
\mathcal Q_c [x] := \lceil x \rceil \,.
$$
Again, for each Monte Carlo run, the noise variance is such that $\frac{\var (Ug) }{\sigma^2} = 10$. We generate $N = 200$ data samples.

\begin{figure}[!ht]
\begin{center}
    {\includegraphics[width=7cm]{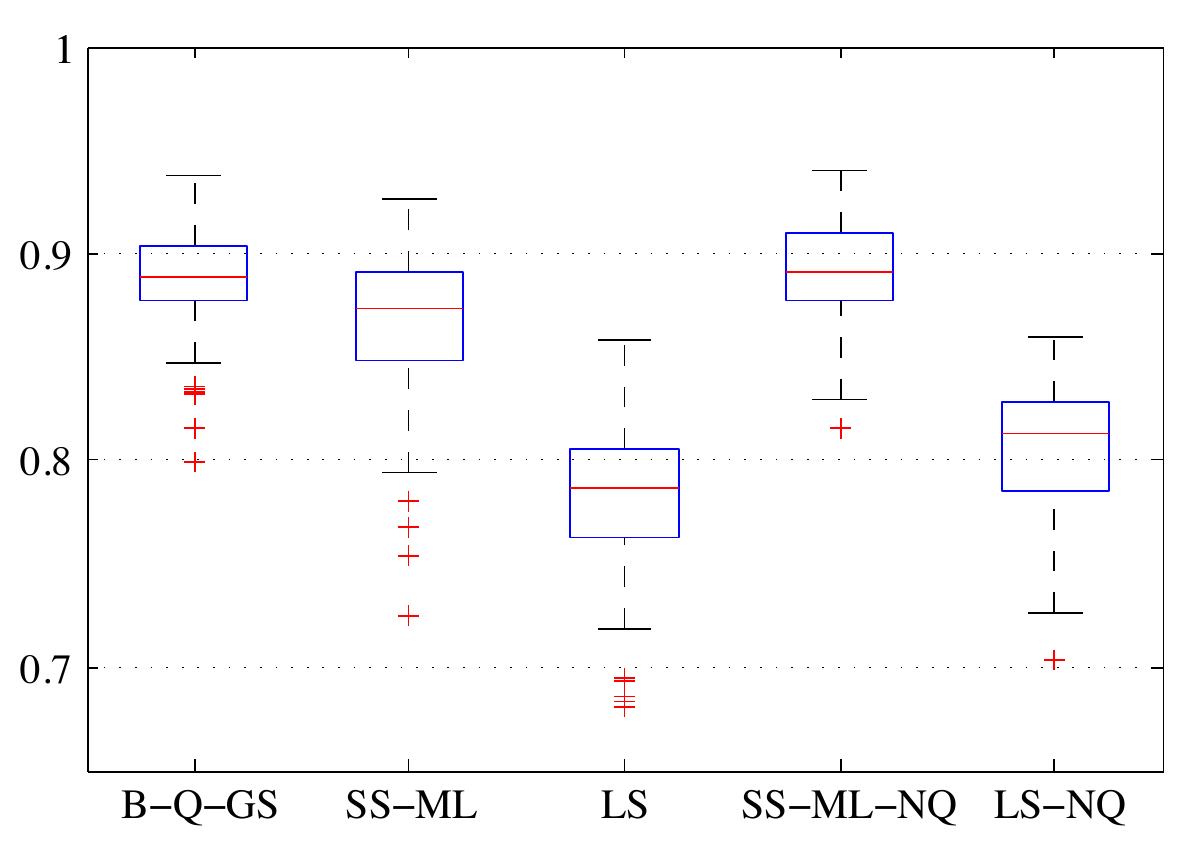}}
\caption{\emph{Box plots of the fitting scores for the ceil-type quantizer experiment.}} \label{fig:ceil_exp}
\end{center}
\end{figure}

As shown in Figure \ref{fig:ceil_exp}, in this case, if one compares the oracle-type methods (i.e. SS-ML-Or. and LS-Or.) with the same methods using quantized data (SS-ML and LS), the loss of accuracy is relatively low. This because this type of quantizer has a mild effect on the measurements. It can be seen, however, that the proposed method is able to give a fit that is comparable to the standard kernel-based method that uses non-quantized data (SS-ML-Or). Moreover, it outperforms the least-squares estimator equipped with the knowledge of non-quantized data.

\section{Conclusions}
In this paper, we have introduced a novel method for system identification when the output is subject to quantization. We have proposed a MCMC scheme that exploits the Bayesian description of the unknown system. In particular, we have shown how to design an integration scheme based on the Gibbs sampler by exploiting the knowledge of the conditional probability density functions of the variables entering the system. We have highlighted, through some numerical experiments, the advantages of employing our method when quantizers affect the accuracy of measurements.

Important questions such as consistency of the method and robust selection of the kernel hyperparameter $\beta$ are currently under study.

\bibliographystyle{plain}

\bibliography{biblio}

\end{document}